\documentclass[prl,aps,amsmath,twocolumn,showpacs,floatfix]{revtex4}
\usepackage{graphicx}
\begin{document}

\title{Penetration of hot electrons through a cold disordered wire. }
\author{A.~S.~Ioselevich and D.~I.~Pikulin}
\affiliation{Landau Institute for Theoretical Physics RAS,  117940  Moscow, Russia,\\
Moscow Institute of Physics and Technology, Moscow 141700, Russia.}
\date{\today}

\begin{abstract}
We study  a penetration of an electron with high energy $E\gg T$ through strongly disordered wire
of length $L\gg a$ ($a$ being the localization length). Such an electron can loose, but not gain
the energy, when hopping from one localized state to another. We have found a distribution function
for the transmission coefficient ${\cal T}$. The typical ${\cal T}$ remains exponentially small in
$L/a$, but with the  decrement, reduced compared to the case of direct elastic tunnelling:
$\overline{\ln{\cal T}}\approx 0.237\cdot 2L/a$. The distribution function has a relatively strong
tail in the domain of anomalously  high ${\cal T}$; the average $\overline{{\cal T}}\propto
(a/L)^{2}$ is controlled by rare configurations of disorder, corresponding to this tail.

\end{abstract}

\maketitle

The electronic transport in disordered one-dimensional systems was extensively studied in the past
50 years \cite{Anderson,review}. In the non-interacting system all the states are localized
\cite{Berezinskii}, so that the transmission
 coefficient ${\cal T}$ of
a finite system exponentially decays with $L$.  ${\cal T}$ varies from sample to sample, since it
depends on the configuration of disorder. For strongly disordered chains the distribution of
$s\equiv -(1/\alpha)\ln {\cal T}$ is a narrow gaussian:
\begin{eqnarray}
F_0(s)\propto\exp\{-(s-1)^2/(\Delta s)^2\},\quad  \Delta s=B\alpha^{-1/2},\label{gauss1}
\end{eqnarray}
\begin{eqnarray}
\alpha= 2L/a\gg 1,\label{gauss0}
\end{eqnarray}
being the principal large parameter of the theory (the measure of the localization strength),
$B\sim 1$ being  the model-dependent factor \cite{noninteracting}. Thus, the conductivity of a
noninteracting system  is zero at $L\to\infty$.

The interactions (e.g., with phonons) lead to a finite equilibrium conductivity $\sigma$ of the
hopping type (see \cite{EfrosShklovskii}). At some $L\sim L_0(T)$ the exponential $L$-dependence of
the conductance $G$ is changed to $G=\sigma/L$, with
 $\sigma$, exponentially
dependent on the temperature $T$ of the system. The specifics of strongly disordered 1d systems
was properly taken into account in
 \cite{Kurkijarvi,RaikhRuzin};
 it was shown that the conductivity in the variable range
 hopping regime is controlled by rare fluctuations of the density of states at the fermi level --
the ``breaks''; as a result
\begin{eqnarray}
\sigma\approx \exp\{-T_0/2T\},\quad T_0=1/ga, \label{gauss2}
\end{eqnarray}
where $g$ is the average density of states. The result \eqref{gauss2} does not obey the Mott law
$\sigma\propto \exp\{-c(T_0/T)^{1/(d+1)}\}$, valid in dimensions $d\geq 2$ (see
\cite{Mott,EfrosShklovskii}), where electrons can easily circumvent the breaks. The
 transport in the strong field ${\cal E}$ was studied in
\cite{Nattermannetal,FoglerKelley,RodinFogler,NguyenShklovskii}. Here the current  $I\propto
\exp\{-8T_0/e{\cal E}a\}$ strongly depends on  ${\cal E}$, not on $T$. It is insensitive to the
breaks, but, on the other hand,  the distribution of electrons is
 far from equilibrium.

In the present paper we study a different situation, where the current through the system arises
due to a {\it small group} of strongly nonequilibrium high-energy particles, so that the occupation
numbers of most electronic states remain essentially in equilibrium.

\begin{figure}
\includegraphics[width=0.9\columnwidth]{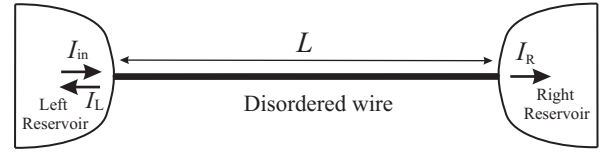}
\caption{The setup. A disordered wire of length $L$ is in a good contact with two reservoirs, the
entire system is in equilibrium. A small flux $I_{\rm in}$ of electrons with energy $E\gg T$ is
sent to the left end of the wire through the left reservoir, and the part $I_R={\cal T}I_{\rm in}$
finally reaches the right reservoir.} \label{setup0}
\end{figure}

 A
disordered wire of length $L$ is in equilibrium with two metallic leads (reservoirs) at temperature
$T$ and chemical potential $\epsilon_F$ (see Fig.\ref{setup0}). In the left reservoir, however, a
small amount of nonequilibrium particles with energies $E\gg T$ ($E$ is measured with respect
 to $\epsilon_F$) is injected, so that the current $I_{\rm in}$
 reaches the left end of the wire.

Hot electrons, injected into the wire, weakly interact with the thermal bath, their energy is not
conserved. However, as long as $E\gg T$, only the processes in which the energy is transferred from
electron to the bath, not vice versa, are allowed. This is only true for not very long wires
$L<L_{\rm Mott}(T)$, where  $L_{\rm Mott}(T)\sim a(T_0/T)^{1/2}$ is the  length of the typical Mott
hop at given $T$. Under this condition  the equilibration does not have chance to occur before the
electrons escape from the wire. In this letter we also do not take into account correlation effects
(like Coulomb gap) due to electron-electron interactions.

Obviously, only the localized states with the energies $\varepsilon_i$ in the interval between the
Fermi energy and the initial energy $E$ of the injected electron are relevant for our problem. We
enumerate them according to their energies: $ 0<\varepsilon_1<\varepsilon_2<\ldots<\varepsilon_n<E.
$ These ``quasiresonant'' levels play an important role in the transport physics, as the electron,
travelling through the chain with initial energy $E$, can make intermediate stops only at these
sites. Indeed, all the sites with $\epsilon_i<0$ are occupied, while the sites with $\epsilon_i>E$
cannot be reached, as no energy can be absorbed. The spatial positions of quasiresonances are
$\ell_i=Lx_i$, the  independent random variables $x_i$ are homogeneously distributed in an interval
$0<x_i<1$. The number $n$ of quasiresonances is itself a random variable, described by the
poissonian distribution $p(n,N)=N^n\exp(-N)/n!$ , where $ N=LEg$ is the average number of
quasiresonances. Thus, each wire is characterized by ``the configuration'' ${\cal C}\equiv\{
n,\{x_1,\ldots x_n\}\}$ and the average of any ${\cal C}$-dependent quantity $A({\cal C})$ over the
ensemble of wires is $ \overline{A}=\sum_{n=0}^{\infty}p(n,N)\prod_{i=1}^{n}\int_0^1dx_iA({\cal
C})$. In particular, the distribution function $F_{N}(s)=\overline{\delta(s-s({\cal C}))}$. In this
letter we focus on the most interesting case $N\gg 1$, when the poissonian distribution is sharp
and one can simply average over $x_i$ at fixed $n\approx N$.

\begin{figure}
\includegraphics[width=0.8\columnwidth]{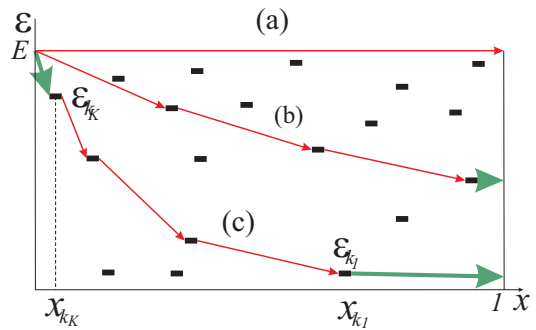}

\caption{(color online) Possible scenarios
 for electron passing from the left to the right reservoir.
``Natural hops'' (occurring with the probability close to unity) are shown by thick green arrows,
the non-natural hops (that happen with exponentially small probability) are shown by thin red
arrows. (a): Direct elastic tunnelling; (b,c): Inelastic staircases with intermediate stops at
quasiresonances. } \label{scenarios}
\end{figure}

How can an electron get from left reservoir to the right one? Besides the obvious possibility of
the direct elastic tunnelling (Fig.\ref{scenarios}a), there are also numerous ``inelastic
staircases''  (Fig.\ref{scenarios}b,c), in which an electrons makes intermediate stops at certain
quasiresonant states, while the excess energy at each hop is transferred to the thermostat. Each
staircase ${\cal S}$ is characterized by the choice of a subset of $K$ ($0\leq K\leq n$)
quasiresonances, with $ \varepsilon_{k_1}<\varepsilon_{k_2}<\cdots<\varepsilon_{k_K}$ and
$x_{k_1}>x_{k_2}>\cdots>x_{k_K}$.

Each staircase contributes to the transmission:
\begin{eqnarray}
{\cal T}({\cal C})=\sum_{{\cal S}}{\cal T}({\cal S}|{\cal C})\propto e^{-\alpha s({\cal C})},\quad
{\cal T}({\cal S}|{\cal C})\propto e^{-\alpha s({\cal S}|{\cal C})},
 \label{cony10rr}
\end{eqnarray}
where the summation runs over all the staircases, possible for given configuration ${\cal C}$.
Under the condition \eqref{gauss0} the sum in \eqref{cony10rr} is dominated by only one -- the
optimal -- staircase ${\cal S}_{\rm opt}({\cal C})$ that corresponds to minimal $s({\cal S}|{\cal
C})$, so that $ s({\cal C})\approx s[{\cal S}_{\rm opt}({\cal C})]=\min_{\cal S}s({\cal S}|{\cal
C})$. In a typical situation the longest hop in the optimal staircase is the last one, then goes
the last but one, etc. Therefore the value of $s({\cal S}|{\cal C})$ is controlled by  few last
hops in ${\cal S}$, while the multitude of short hops in the upper part of the staircase are of
only secondary importance. The most natural assumption about the  structure of the optimal
staircase would be {\it the scaling hypothesis}: the distribution function ${\cal P}_k(\ell_k)$ for
random variables $\ell_k=x_{i_k}/x_{i_{k-1}}$ does not depend on $k$. Such a simple self-similar
structure was, however, not observed in our numerical experiments: ${\cal P}_k$ manifestly depended
on $k$.

The explicit expression for $s({\cal S}|{\cal C})$ can be found with the help of the stationary
master equation for the quasiresonant levels populations $f_i$:
\begin{eqnarray}
P_{L\to i}I_{\rm in}+\sum_{j>i}f_jP_{j\to i}-f_iP_{i\to {\rm out}}=0, \label{cony1}
\end{eqnarray}
The first term in \eqref{cony1} is the incoming flux of particles from the left reservoir; the
second term describes the particles, coming to the level $i$ from all other levels with higher
energies (hence the condition $j>i$); finally, the third term takes into account all possible
escapes from the $i$-th level: $P_{i\to {\rm out}}=P_{i \to L}+P_{i \to R}+\sum_{j<i}P_{i\to j}$.
The system's transmittance is
\begin{eqnarray}
{\cal T}\equiv I_R/I_{\rm in}=P_{L \to R}+\sum_{i}P_{i \to R}(f_i/I_{\rm in}). \label{cony1a}
\end{eqnarray}
 The rate of transitions between $j\to i$ is
$P_{j\to i}=P^{(0)}_{j\to i}(\varepsilon_j-\varepsilon_i)e^{-\alpha |x_i-x_j|}$. Matrix elements of
the electron-thermostat interaction, entering $P^{(0)}_{j\to i}$ are smooth power-law functions of
the energy transfer $\varepsilon_j-\varepsilon_i$. It means, that if we are interested only in the
exponential dependencies, we do not have to take these matrix elements into account. Thus, in the
exponential approximation we can write
\begin{eqnarray}
 P_{i\to j}\propto\theta(i-j)e^{-\alpha|x_i-x_j|},\quad
P_{i\to {\rm out}}\propto e^{-\alpha \chi_i},\\
 \chi_i({\cal C})\equiv
\min\left\{x_i,1-x_i,\min_{j<i}\{|x_i-x_j|\}\right\},
 \label{cony4}
\end{eqnarray}
being the distance from the $i$-th quasiresonance to its ``natural descendant'' -- a closest
neighbor with lower energy,  or to one of the two reservoirs. The solution of the system of
equations \eqref{cony1} can be written in a recurrent form:
\begin{eqnarray}
f_i=(P_{L\to i}/P_{i\to {\rm out}})I_{\rm in}+\sum_{j>i}f_j(P_{j\to i}/P_{i\to {\rm out}}),
\label{cony4uu}
\end{eqnarray}
which allows for finding $f_i$ provided all $f_j$ with $j>i$ are already found. According to the
exponential approximation, justified by the large parameter \eqref{gauss0}, any sum, occurring in
\eqref{cony1a} or in \eqref{cony4uu}, is dominated by a single term with the smallest negative
exponent. As a result, the normalized probability for the electron to make a hop $i\to j$ is
$p(i\to j)=P_{i\to j}/P_{i\to {\rm out}}=\exp\{-\alpha \Delta s(i\to j)\}$, where $\Delta s(i\to
j)=|x_{i}-x_{j}|-\chi_{i}$. Then, having in mind that ${\cal T}({\cal S}|{\cal
C})=\prod_{k=1}^{K+1} p(i_k\to i_{k-1})$, we arrive at
\begin{eqnarray}
s({\cal S}|{\cal C})=\sum_{k=1}^{K+1}\Delta s(i_k\to i_{k-1}),
 \label{cony7r}
\end{eqnarray}
so that $\Delta s(i_k\to i_{k-1})$ has the meaning of the contribution of the $k$-th hop to the
exponent $s$. In \eqref{cony7r} $x_{0}=1$ and $x_{K+1}=0$ are the positions of the right and the
left reservoirs, correspondingly; $\chi_{K+1}=0$. From \eqref{cony7r} it immediately follows that
$s({\cal S}|{\cal C})=1-\sum_{k=1}^{K+1}\chi_{i_{k}}$. In principle, our problem can be solved by
means of enumeration of all staircases, possible for a given configuration, and choosing the
optimal one. The direct solution of the master equations, however, leads to the same result.

Note, that $\Delta s(i_k\to i_{k-1})=0$ if the $i_{k-1}$-th quasiresonance is the natural
descendant of the $i_k$-th one. The corresponding hops we will call ``natural hops'' in what
follows. Clearly, to minimize $s$, it would be nice to have a staircase, where all  the hops
$i_k\to i_{k-1}$ (or at least as many of them, as possible) are natural. We will see, however, that
such a ``natural staircase'' is possible to find only for some  rare ``fortunate configurations''.

\begin{figure}
\includegraphics[width=0.8\columnwidth]{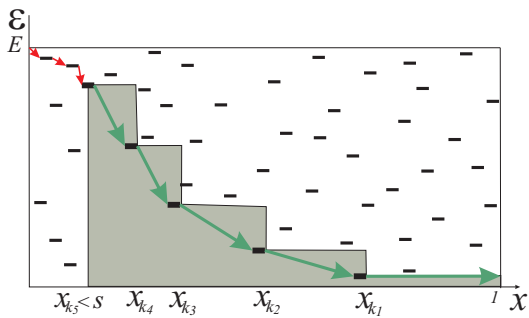}

\caption{(color online) A ``fortunate'' configuration (with $m=5$).  The last $m$  hops in an
optimal staircase are ``natural''. There are no quasiresonances in the shaded domain.}
\label{fortunate}
\end{figure}

Suppose that certain configuration ${\cal C}$
 generates  an optimal staircase, in which a sufficiently long subsequence
of  last $m$ hops  ${k_m}\to{k_{m-1}}\to\cdots\to{k_1}\to R$ is ``natural'', see
Fig.\ref{fortunate}. It means that, if an electron somehow manages to get to the upper level
${k_m}$ in this subsequence, then it makes its descending way  to the right reservoir through the
rest of the staircase with a probability that is close to unity. Then, if this upper level is close
enough to the left reservoir (namely, if $x_{k_m}<s$), then one can guarantee that $s({\cal C})<s$,
no mater whether the preceding short hops in the staircase are natural or not.
The necessary conditions  for a given configuration ${\cal C}$ to be a fortunate one are as
follows:
 The last jump (to $R$) should start from the lowest level (i.e., $k_1=1$), and
this level should be localized in the right half of the wire: $x_1>1/2$.
 In general, the level $k_p$ in the optimal staircase should be the lowest one among
all the quasiresonances with $0<x_i<x_{k_{p-1}}$; this level should be in the right half of the
stretch: $x_{k_{p-1}}/2<x_{k_{p}}<x_{k_{p-1}}$. This should be true for all jumps with $p<m$, where
$m$ is determined by the condition $x_{k_m}<s$.

 The probability for all the $m$ last jumps in the optimal staircase to
 be natural is $2^{-m}$.
 However, $m$
 is not determined solely by $s$: for fixed $s$ $m$ still fluctuates from
 configuration to configuration.

 Let us introduce statistically independent random variables $\xi_p=2[x_{k_{p}}-x_{k_{p-1}}]/x_{k_{p-1}}$,  homogeneously distributed in the interval
 $0<\xi_p<1$. Then
$s=\prod_{p=1}^{m}(1+\xi_p)/2$. The distribution $P_m(\phi)$ of the random variable $\phi=m\ln
2-\ln(1/s)=\sum_{p=0}^m\ln(1+\xi_p)$ can be obtained with the help of the Fourier transformation:
\begin{eqnarray}
P_m(\phi)=\int\frac{dq}{2\pi}e^{-iq\phi}\left[\frac{2^{iq+1}-1}{iq+1}\right]^m, \nonumber
\end{eqnarray}
\begin{eqnarray}
F_{\infty}(s)\approx\sum_{m}2^{-m}\int d\phi P_m(\phi)\delta(s-\exp\{\phi-m\ln 2\})=\nonumber\\
=\sum_{m}\frac{P_m(m\ln 2-\ln(1/s))}{s2^{m}}
=\int\frac{dq}{2\pi s}\frac{(1+iq)e^{iq\ln(1/s)}}{2^{-(1+iq)}+iq}\nonumber\\
\propto s^{-1}e^{-2\ln(1/s)}\approx bs.\nonumber
\end{eqnarray}
where $b\sim 1$ is a universal constant. Although this asymptotics describes only a small fraction
of the configurations, it turns out to be sufficient for finding the average transmission
coefficient:
\begin{eqnarray}
\overline{\mathcal T}\approx\int_0^1 bs \exp\{-\alpha s\} ds=b\alpha^{-2}\propto L^{-2}.\label{4xe}
\end{eqnarray}
\begin{figure}
\includegraphics[width=0.9\columnwidth]{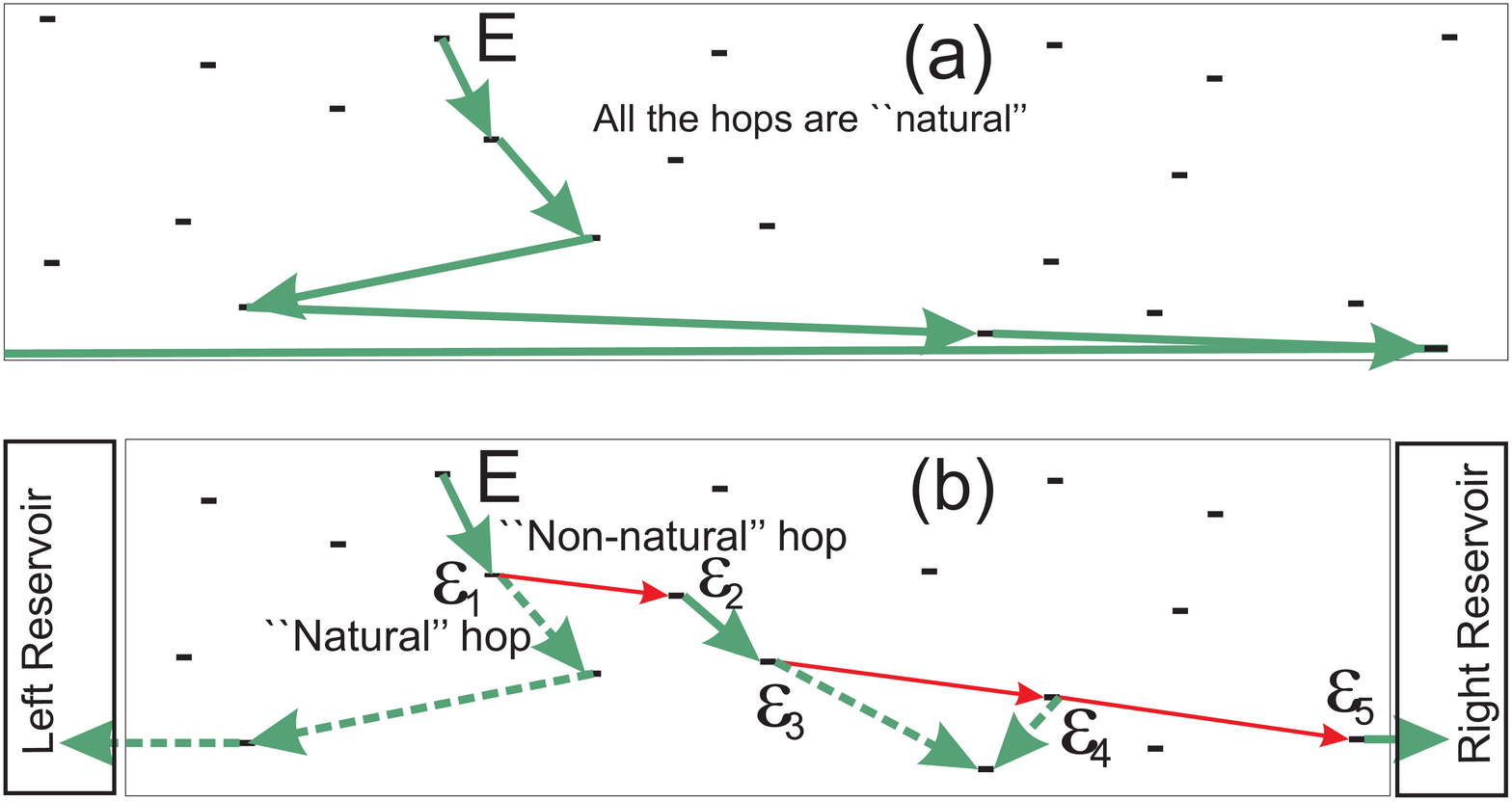}
\caption{(color online) (a): The ``natural path'' consisting of hops to nearest neighbors with
lower energy.  The particle starts from the middle of the wire, so that in one half of the
ensemble of samples the natural path leads the left reservoir, and it leads to the right one in the
other half. (b): The particle is initially placed near the left reservoir, the natural path leads
there in almost all samples. To get to the right reservoir, a particle has to make non-natural
hops.}\label{diffusion}
\end{figure}

The hopping motion of hot particles, accompanied by the emission of energy, earlier was  studied
in the context of recombination of photo-excited  electron-hole pairs \cite{Shklovskii}. For that
end it was sufficient to take into account {\it only the natural hops} since a particle was assumed
to be created in the bulk, far from the ends of the sample, so that it could not  escape to a
nearby lead (see Fig.\ref{diffusion}a). Under these conditions the evolution of particle density
profile, {\it averaged  over the  ensemble of samples} could be described by a peculiar diffusion,
in which the distribution  of  hops lengths is rescaled by a fixed parameter $q>1$ after each hop.
If one would extend the same approach to the setup where the particle is initially placed near one
of the absorbing leads, then one easily finds for the probability to reach the opposite lead (and
thus the average transmission coefficient)  $\overline{\mathcal T}\propto L^{-\beta}$, where the
exponent $\beta$ (as well as the parameter $q$) is universal. Note that this result is in agreement
with \eqref{4xe}. However, as we have already mentioned, $\overline{\mathcal T}$ is controlled by
very rare anomalous samples with high transparency, while in a typical sample the natural path
would lead to the nearby left lead (see Fig.\ref{diffusion}b). Therefore, to find the probability
to reach the right lead in a typical sample one should take into account the non-natural hops, that
were ignored in \cite{Shklovskii}. The ``diffusional approach'' \cite{Shklovskii}, being an
adequate instrument for finding $\overline{\mathcal T}$, is useless for the determination of the
distribution of ${\mathcal T}$.

\begin{figure}
\includegraphics[width=0.8\columnwidth]{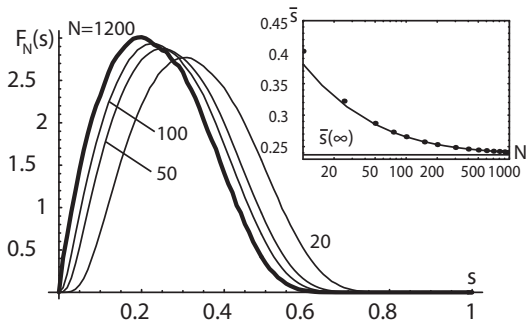}

\caption{Distribution functions of the logarithm of transparency
 for different values of average
number  of quasiresonaces $N\equiv ELg$. For $N\gtrsim 200$ the curves practically merge and
coincide with $F_{\infty}(s)$, shown by thick line. The $N$-dependence of the average logarithm of
transmittance
 is shown on the inset, the solid line being the asymptotic expression \eqref{cony11ereq}.} \label{distr}
\end{figure}

The distribution $F_{N}(s)$ for  general $s$ and $N$ can only be found numerically. We generated an
ensemble of $\sim 10^6$ random configurations ${\cal C}$ and calculated corresponding $s({\cal C})$
with the help of the recurrent formula \eqref{cony4uu}. The results  of our Monte-Carlo simulations
are summarized in Fig.\ref{distr}. The distribution functions $F_{N}(s)$ are wide: the dispersion
of $s$ is of order of $\overline{s}$ for all $N$. For $N\gtrsim 300$  practically $F_N(s)\approx
F_{\infty}(s)$, and $F_{\infty}(s)$ indeed shows a linear low-$s$ asymptotics with $b\approx 29$.

 The average value $\overline{s}(N)$  monotonically decreases
 with $N$,  tending to a finite limit
$\overline{s}(\infty)\approx 0.237$ at $N\to \infty$. The convergence is, however, extremely slow:
for $N=1000$ the relative difference is still of order of $2\%$. The convergence can be improved
dramatically if one introduces an asymptotic correction according to the empiric formula
\begin{eqnarray}
\overline{s}(N)\approx 0.237+0.598\ln N/N,\qquad \mbox{for $N\gg 1$}.
 \label{cony11ereq}
\end{eqnarray}
The deviation of experimental $\overline{s}(N)$ from the asymptotics \eqref{cony11ereq} is less
than $2\%$ already for $N\sim 30$. The $N$-dependent corrections are controlled by many short hops
in the beginning of the optimal staircase. The scaling hypothesis, mentioned above, would lead to
$\overline{s}(N)-\overline{s}(\infty)\propto 1/N$, which is inconsistent with our numerical data.
This is another argument against the scaling.  A fit of our numerical data for the $N$-dependence
of an average number $\overline{K}$ of hops in the optimal staircase gives
\begin{eqnarray}
\overline{K}(N)\approx 0.39(\ln N)^2+2.4,\qquad \mbox{for $N\gg 1$}
 \label{nfitq1}
\end{eqnarray}
which is perfectly consistent with the result (\ref{cony11ereq}) and, again, inconsistent with the
scaling hypothesis, (the latter would give $\overline{K}(N)\propto\ln N$). The elucidation of the
structure of the initial part of the optimal staircase and the origin of the empirical laws
(\ref{cony11ereq},\ref{nfitq1})  still  remains a challenge.

In conclusion, we have found the distribution function of the transmission coefficient for the
inelastic penetration of a cold disordered wire by a hot electron. The applications of these
results to  specific physical effects will be presented in a long paper to follow.

We are indebted to M.E.Raikh for valuable comments.

\end{document}